\documentclass[universe,article,accept,pdftex,moreauthors]{Definitions/mdpi} 
\firstpage{1} 
\pubvolume{1}
\issuenum{1}
\articlenumber{0}
\pubyear{2026}
\copyrightyear{2026}
\externaleditor{~} 
\datereceived{16 June 2026} 
\daterevised{29 July 2026} 
\dateaccepted{27 August 2026} 
\datepublished{ } 

\Title{Spheroidal Resampling Analysis of High-Redshift Gamma-Ray Burst Spatial Densities}

\Author{Istvan Horvath
 $^{1,}$*\orcidA{}, Zsolt Bagoly $^{1,2}$\orcidB{}, Lajos G. Balazs
 $^{3,4}$\orcidC, Jon Hakkila $^{5}$\orcidD, Janos Horvath $^{6}$\orcidE{}, Sandor Pinter $^{1}$\orcidF, Istvan I. Racz $^{1}$\orcidG, Peter Veres $^{7,8}$\orcidV ~and Bendegúz Koncz $^{4,9}$\orcidH{}}

\AuthorNames{Istvan Horvath, Zsolt Bagoly, Lajos G. Balazs, Jon Hakkila,  Janos Horvath, Sandor Pinter, Istvan I. Racz, Peter Veres, Bendegúz Koncz}

\address{%
$^{1}$ \quad Department of Natural Science, University of Public Service, H-1101, Budapest,
 Hungary; \linebreak  bagoly.zsolt@uni-nke.hu (Z.B.); sandor.pinter@uni-nke.hu (S.P.);
 racz.istvan@uni-nke.hu (I.I.R.)  \\
$^{2}$ \quad Department of Physics of Complex Systems, E\"otv\"os University, H-1117, Budapest, Hungary \\
$^{3}$ \quad Konkoly Observatory, Research center for Astronomy and Earth Sciences, H-1121, Budapest, Hungary; balazs@konkoly.hu\\
$^{4}$ \quad Department of Astronomy, E\"otv\"os Lor\'and University, H-1117, Budapest, Hungary; kbendeg398@mailbox.unideb.hu, lgbalazs@staff.elte.hu  \\
$^{5}$ \quad Department of Physics and Astronomy, The University of Alabama in Huntsville, Huntsville, AL 35899, USA; jh0271@uah.edu  \\
$^{6}$ \quad Visionary Tech \& Event Solutions, Sacramento, CA 95816, USA; horvath\_janos@visionarytecheventsolutions.com\\
$^{7}$ \quad Department of Space Science, University of Alabama in Huntsville, Huntsville, AL 35899, USA; pv0004@uah.edu\\
$^{8}$ \quad Center for Space Plasma and Aeronomic Research, University of Alabama in Huntsville, \linebreak  Huntsville, AL 35899, USA\\
$^{9}$ \quad Faculty of Science and Technology, University of Debrecen,
 4032 Debrecen, Hungary}

\corres{Correspondence: horvath.istvan@uni-nke.hu}

\abstract{Gamma-ray bursts (GRBs) are bright transient sources that can be observed at high redshift. They can therefore be used as tracers of the distant large-scale structure, although the GRB redshift sample is sparse and affected by strong selection effects. We analyze the three-dimensional distribution of 542 GRBs with spectroscopic redshifts. The method extends our earlier spherical-window search by replacing spherical counting volumes with axisymmetric spheroids. The angular positions of the observed GRBs are kept fixed in the Monte-Carlo null samples, while the redshifts are shuffled within each Galactic hemisphere. 
In the Northern Galactic hemisphere, the overdensity associated with the Hercules--Corona Borealis Great Wall remains significant for a range of spheroidal shapes. This supports the stability of the previously reported signal and shows that it is not only a consequence of using spherical counting volumes. In the Southern Galactic hemisphere, the same method finds no structure with comparable significance. A small candidate grouping is present, but its significance is only marginal, and it is sensitive to the small number of events. We conclude that spheroidal resampling is a useful check of GRB overdensity searches, but the physical nature of any candidate structure still requires confirmation with independent tracers such as galaxy or quasar samples.}

\keyword{\textls[-15]{large-scale structure of the universe; gamma-ray bursts; cosmology;}
 observations; methods; statistical; methods; data analysis}

\newcommand{\apj}{ApJ}
\newcommand{\apjl}{ApJL}

\newcommand{\nat}{Nature}

\begin{document}

\section{Introduction}
\label{sec:introduction}

Gamma-ray bursts (GRBs) are short flashes of gamma-ray radiation followed, in~many cases, by~afterglows at longer wavelengths (e.g.,~\citep[]{mesz06,2016SSRv..202..195P,zhang_2018}). Long-duration GRBs are linked mainly to the collapse of massive stars, while short-duration GRBs are linked mainly to compact-object mergers~\citep{galaxies10030066,1999ApJ...524..262M,berger14,2017ApJ...848L..13A}. Because~GRBs are very luminous, they can be detected at large cosmological distances
~\citep{2006Natur.443..186I, Salvaterra_2009, 2015A&A...581A..86M}. When an afterglow is observed, its sky position can be measured accurately, and~a spectroscopic redshift can sometimes be obtained~\citep{Wang15,2016SSRv..202..195P}. Long-duration GRBs are therefore useful, although~sparse, {as tracers of massive star formation and of the distant matter distribution~\citep{2012ApJ...755...85M, Kistler_2009}}. 
{Large-scale structures have also been found in galaxy and quasar surveys, providing an important comparison for GRB-based studies}~\mbox{\citep{2019MNRAS.490.4481And, Gott05,clo12,2021MNRAS.507.1361M,2017JCAP...06..019N}.}

On very large scales, the~standard cosmological model assumes that matter is statistically homogeneous and isotropic. This assumption is usually called the Cosmological Principle~\citep{2000ApJ...536....1L,Shawqi_Al_Dallal_2024}.
Recent reviews have summarized several observational tensions with statistical isotropy and homogeneity, emphasizing that claims of departures from the Cosmological Principle require careful treatment of tracer selection effects and survey systematics~\citep{2023CQGra..40i4001K}. On~smaller and intermediate scales, matter forms the cosmic web of filaments, walls, clusters, and~voids~\citep{1996Natur.380..603B, 2005Natur.435..629S}. A~key observational question is the scale at which these structures average out. Large galaxy and quasar surveys are the main tools for testing this transition to homogeneity~\citep{Gott05,clo12,2020ApJ...897..133P}. At~high redshift, however, the~available tracers change. GRBs offer an additional way to test whether very large structures can be seen in the distant Universe~\citep{2012MNRAS.419..556C,2022MNRASLopez,2024JCAP...07..055L}.

Several large structures have been proposed from GRB samples. The~Hercules--Corona Borealis Great Wall is the most discussed and statistically-significant GRB cluster. It was reported as an excess of GRBs in the Northern Galactic hemisphere and has an estimated size of about 2--3 Gpc~\citep{hhb14,hbht15,HSZ20}. Another cluster is the Giant GRB Ring, a~ring-like distribution of GRBs in the redshift range $0.78<z<0.86$, with~a diameter of about 1.7 Gpc~ \mbox{\citep{BalazsRing2015,BalazsTus2018}. }
These clusters are important candidate cosmological structures, but~interpretation of them must be made carefully due to the presence of strong observational selection biases. The~GRB redshift sample is incomplete, being shaped by satellite triggering, sky exposure, afterglow brightness, dust extinction, telescope scheduling, and~the chance of obtaining a spectrum~\citep{2011A&A...526A..30G,Schulze15}. However, by~accounting for these biases, we seek to demonstrate that the observed sample is adequate for use in large-scale structure~studies.

Observational selection effects affect all areas of astronomy. However, due to the transient, high-energy nature of GRB emission, and~to the natural delay occurring between this emission and the optical/infrared afterglow, GRBs are especially susceptible to selection biases. For~example, the~\textit{Swift}
 Burst Alert Telescope (which was designed specifically for detecting GRBs and their associated afterglows) has a non-uniform sky exposure due to the pointing history of the {satellite and the sensitivity of the detector}~\citep{2004ApJ...611.1005G,2005SSRv..120..143B,2012A&A548L7Tello,2006A&A...447..897J}. 

A detailed map of this exposure was derived by \citet{2025ApJ989.161Lien}. Recent methods, such as a factorized Bayesian model based on Spherical Generalized Additive Models, can separate wideband exposure from narrowband redshift-selection effects~\citep{2026Univ...12...31B}. Other biases may come from the way follow-up time is assigned by observers, because~the chance of measuring a redshift can depend on target selection and telescope scheduling~\citep{2022Univ....8..342B,2025AcPol..65....9B}. Earlier studies of GRB directional anisotropy in prompt-emission properties did not find strong evidence for similar large-scale directional effects~\citep{2017ApJ...851...15R,2019MNRAS.486.3027R,2022Univ....8..342B}. In~the present paper we take a simpler and more direct approach to accounting for these known biases: We keep the observed angular positions fixed and randomize only the redshift information. This means that the Monte-Carlo samples keep the observed sky footprint and its main angular~biases.

This work extends the analysis of \citet{horv26MNRAS}, hereafter Paper~I. 
Paper~I searched for overdensities using spherical counting volumes in a transformed three-dimensional coordinate system. Here we use the same GRB sample and the same redshift-shuffling model, but~we replace spheres by axisymmetric spheroids. The~aim is to test whether the Northern overdensity remains significant when the shape of the counting volume is changed. We also test whether a similar signal appears in the Southern Galactic hemisphere. In~this paper, any overdensity found from the GRB redshift sample alone is treated as a candidate structure. Its physical nature must be checked with independent tracers in the same comoving~volume.

The paper is organized as follows. Section~\ref{sec:methodology} describes the data, the~spheroidal counting volumes, and~the Monte-Carlo test. Sections~\ref{sec:north} and \ref{sec:south} give the results for the Northern and Southern Galactic hemispheres, respectively. Section~\ref{sec:discussion} discusses the interpretation and the main limits of the method. Section~\ref{sec:summary} summarizes the~results.

\section{Methodology} 

\label{sec:methodology}

This section describes the GRB sample, the~coordinate transformation, the~spheroidal counting volumes, and~the Monte-Carlo resampling used to estimate the statistical significance of the overdensities~\citep{horv24MNRAS}.

\subsection{GRB~Dataset}
\label{sec:grbdataset}

We use the same GRB sample as in Paper~I~\citep{horv26MNRAS}. The~sample contains 542 GRBs with reliable spectroscopic redshifts and measured sky positions. The~data were collected from public GRB databases, mainly the GRBOX database and the table compiled by Jochen Greiner. We also used Gamma-ray Coordinates Network circulars when needed. The~data sources are listed in the Data Availability Statement.
The sample is a spectroscopic-redshift sample, not a complete GRB sample. It includes both long- and short-duration GRBs, depending on the objects with secure redshifts in the input~catalog.

\begin{figure}[H]

\includegraphics[width=0.98\columnwidth]{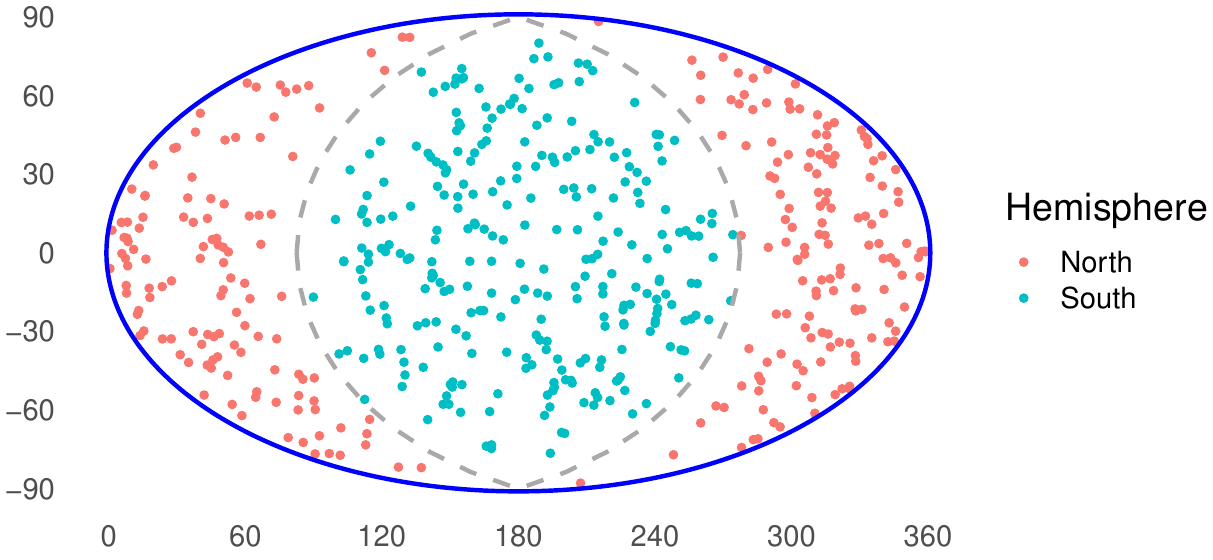}%
\caption{Sky
 distribution of the 542 GRBs, shown as a projection viewed from the direction of the Southern Galactic Pole (center of plot).
The dashed grey curve marks the Galactic equator.}
\label{fig:south_sky_paper1}
\end{figure}

The sample is artificially divided at the Galactic equator. The~Northern Galactic hemisphere, defined by $b>0^{\circ}$, contains 262 GRBs. The~Southern Galactic hemisphere, defined by $b<0^{\circ}$, contains 280 GRBs. The~sky distribution of the full sample is shown in Figure~\ref{fig:south_sky_paper1}. Paper~I showed that the two hemispheric samples have similar observational distributions in~redshift.

\unskip

\subsection{Analysis~Method}
\label{sec:method}

The method follows Paper~I, with~one change: the counting volumes are no longer restricted to spheres. We use axisymmetric spheroids to test whether a detected overdensity depends on the assumed shape of the counting~volume.

\subsubsection{Radial~Transformation}

{If the GRB distribution were homogeneous, the~cumulative number of objects enclosed within a sphere of radius $r$ would increase as $r^3$. The~observed GRB sample does not follow this relation in its original radial coordinate: the number density decreases systematically with distance because of observational selection effects and the possible evolution of the GRB population. A~radial correction is therefore needed before local three-dimensional densities can be compared.}

{As discussed in Paper~I~\citep{horv26MNRAS}, this radial trend can be treated in two general ways. One possibility is to estimate the local radial density and assign each GRB a weight proportional to the reciprocal of that density; the density within a test volume is then obtained by summing the weights of the enclosed GRBs. The~second possibility is to redefine the radial coordinate so that cumulative object number becomes approximately proportional to the third power of the transformed distance. We use the latter approach in the primary analysis, as~in Paper~I (Figure 2).}

\begin{figure}[H]

\includegraphics[width=0.98\columnwidth]{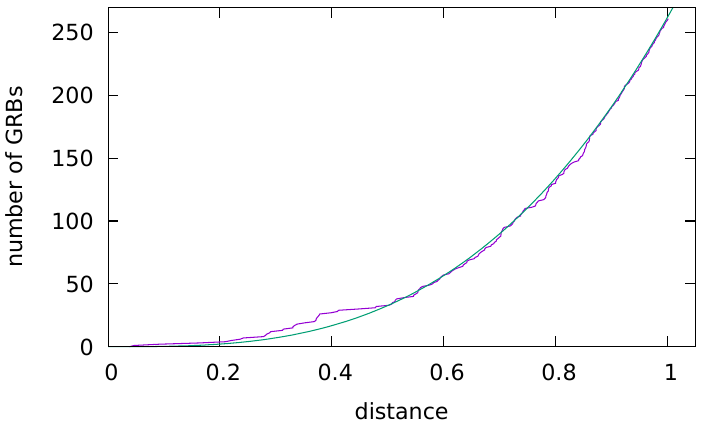}%
\caption{Cumulative
 number of Northern-hemisphere GRBs as a function of transformed radial distance. The~purple curve corresponds to the quasi-cubic radial transformation used in Paper~I and in the primary analysis of this work. The~green curve shows the exact cubic relation, $N(<\rho)=N\rho^3$, expected for a constant spatial number density.}
\label{fig:koboseszaki}
\end{figure}

{A second, exact cubic radial transformation can be defined directly from the ranks of the GRBs. The~objects are ordered according to any radial quantity that increases monotonically with redshift. If~$i=1,\ldots,N$ denotes the rank of an object in a sample of size $N$, its normalized cumulative rank is
 $u_i=i/N$, and~the alternative transformed radius is defined as}
\begin{equation*}
\boldsymbol{\rho_i=\left(\frac{i}{N}\right)^{1/3}}.
\end{equation*}
{It then follows by construction that $N(<\rho)=N\rho^3$. Since only the ordering of the objects is used, this exact rank-based transformation is independent of the particular radial distance definition, provided that it is a monotonically increasing function of redshift.}

{After the radial transformation, the~GRBs are placed in a three-dimensional Cartesian coordinate system. The~centers of the counting volumes are placed on a regular three-dimensional grid with lattice spacing $l=0.01$ in the transformed coordinate units. The~grid covers the full volume occupied by the transformed GRB sample. For~each test center, we count the number of GRBs contained within the selected volume, and~the maximum count over all grid centers is denoted by $K_{\rm obs}$. Further details of the baseline procedure are given in Paper~I.}

\subsubsection{Spheroidal Counting~Volumes}
\begin{figure}[H]
\centering
\includegraphics[width=0.98\columnwidth]{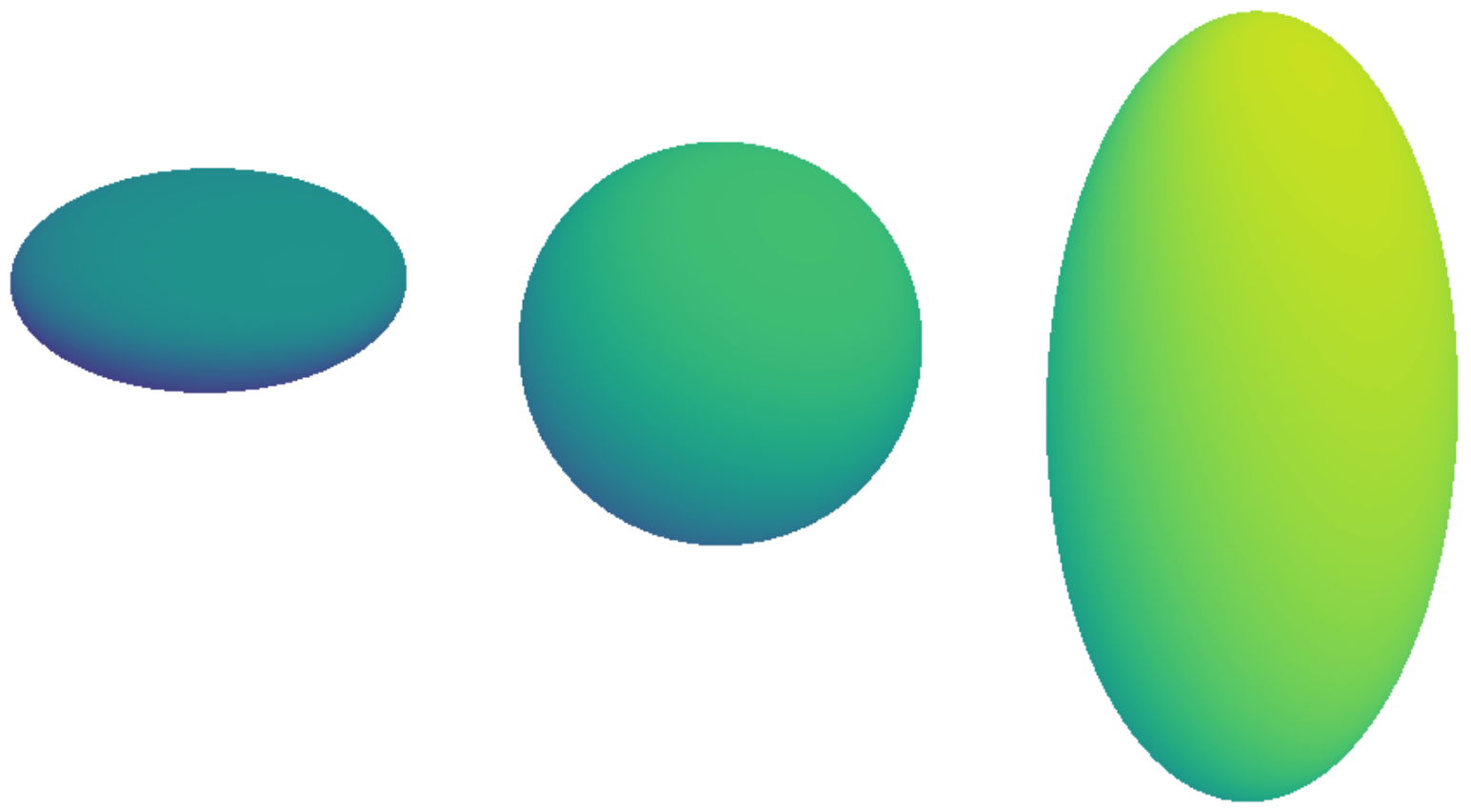}%
\caption{Schematic
 view of the counting volumes. The~spherical case, $s=1$, is shown in the center. Spheroids with $s<1$ are flattened along the $z$ direction, while spheroids with $s>1$ are elongated along the $z$ direction.}
\label{fig:spheroid_schematic}
\end{figure}

For a test center and a GRB, let $(\Delta x,\Delta y,\Delta z)$ be the coordinate differences in the transformed Cartesian system. We define the spheroidal distance by
\begin{equation}
d_s^2 = (\Delta x)^2 + (\Delta y)^2 + \left(\frac{\Delta z}{s}\right)^2 .
\label{eq:sphdist}
\end{equation}
A GRB
 is counted if
\begin{equation}
d_s^2 \le r^2 .
\label{eq:sphinside}
\end{equation}
With this definition, the~counting volume has semi-axes $(r,r,rs)$, {where} $s$ is a dimensionless parameter. The~case $s=1$ is the spherical method, which was already {published in Paper~I}. Values $s>1$ describe volumes elongated along the $z$ direction, while values $s<1$ describe flattened volumes (see Figure~\ref{fig:spheroid_schematic}). The~radius $r$ is the transverse semi-axis of the spheroid, and~the corresponding volume is proportional to $r^3s$.

{We also note that a spherical counting volume ($s = 1$) assumes spatial isotropy. This assumption can introduce a shape bias during the search for large-scale structure candidates. Cosmic overdensities, for~instance, planar walls or filaments, are inherently anisotropic. If~a spherical window is forced upon an elongated or flattened structure, underdense background regions are included. This geometric mismatch results in signal dilution. Furthermore, the~line-of-sight coordinates are derived from redshift measurements and radial distance transformations. Because~of this, the~radial axis is methodologically different from the transverse angular directions. In~standard cosmology, the~radial distance is mathematically equivalent to the look-back time. Consequently, the~radial axis represents the axis of cosmic evolution. The~introduction of the stretching factor $s$ decouples the radial scales from the transverse scales. This formulation allows us to measure the geometrical robustness of the detected structures. It guarantees that the physical signal is not an artificial consequence of an isotropic filtering window.}

\vspace{-3pt}
\unskip

\subsubsection{Monte-Carlo Significance~Test}
The null hypothesis follows Paper~I and the related bootstrap-resampling approach discussed by \citet{horv25Universe}. For~each Galactic hemisphere, we keep the angular positions.~The~redshifts are then randomly permuted among the different observed angular positions in that hemisphere. This produces synthetic catalogs that keep the same sky footprint and the same one-dimensional redshift distribution as the observed sample but~remove any original three-dimensional~clustering.

The present redshift-permutation test and the fully homogeneous-isotropic Monte-Carlo test correspond to different null hypotheses. \citet{nada13} applied the same structure-finding algorithm to homogeneous Poisson catalogues in order to determine how frequently extreme apparent structures arise in an idealized homogeneous point process. In~contrast, our test is conditional on the observed angular and radial marginal distributions. It asks whether the observed association between redshifts and sky positions produces an unusually large three-dimensional concentration. We use this conditional null hypothesis because the spectroscopic-redshift GRB sample has a non-uniform and incompletely modelled angular selection function. Replacing the observed directions with uniformly distributed directions would not preserve these observational effects. Consequently, the~probabilities reported here are conditional permutation probabilities and should not be interpreted as probabilities under a fully homogeneous and isotropic Poisson null model.

For each pair $(r,s)$, the~observed maximum count $K_{\rm obs}(r,s)$ is compared with \linebreak  $W=10000$ \textls[-15]{random catalogs. The~ probability of getting $K_{\rm obs}(r,s)$ is estimated as the fraction of simulations in which the maximum count is at least as large as the observed value,}
\begin{equation}
p(r,s) = \frac{N\left[K_{\rm sim}(r,s) \ge K_{\rm obs}(r,s)\right]}{W} .
\label{eq:pvalue}
\end{equation}
The scan over many values of $r$ and $s$ introduces a look-elsewhere effect, by~which significant results can be artificially obtained by inadvertently overlooking insignificant results. Unless~stated otherwise, the~probabilities quoted below should therefore be read as local probabilities for the tested parameter~values.

\section{The Northern Hemisphere GRB~Distribution}
\label{sec:north}

The Northern Galactic hemisphere contains 262 GRBs. We first check that the method reproduces the spherical result of Paper~I. We then allow $s$ to vary and test whether the Northern overdensity remains significant for non-spherical counting~volumes.

\subsection{Recovery of the Spherical~Result}

\label{sec:north_spherical}

For $s=1$, the~spheroid becomes a sphere and the method is the same as in Paper~I. The~algorithm again finds the main Northern overdensity associated with the Hercules--Corona Borealis Great Wall. The~largest count is $K=125$ GRBs. Depending on the adopted scale, the~local Monte-Carlo probability lies in the range $0.0007 \lesssim p \lesssim 0.01$, as~similarly reported in Paper~I. This recovery is an important check, because~it shows that the present code gives the same baseline result before the spheroidal extension is~applied.

\subsection{Dependence on Spheroidal~Shape}

\label{sec:north_shape}

\begin{figure}[H]

    \includegraphics[width=0.97\columnwidth]{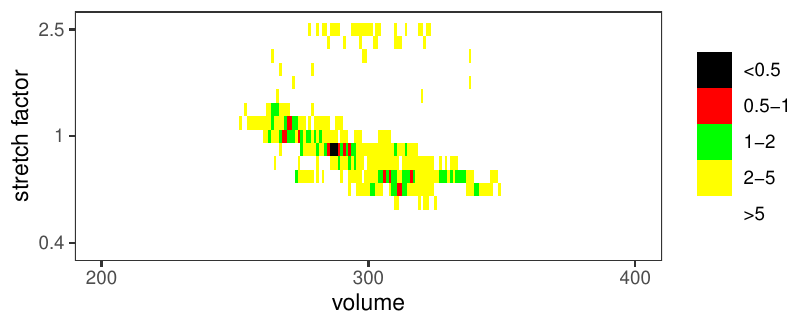}
    \caption{
    Monte-Carlo probabilities for the Northern Galactic hemisphere. The~horizontal axis shows the tested volume parameter, and~the vertical axis shows the stretching factor $s$. The~color bar gives the Monte-Carlo probability $p$ in percent. Only the range $200 \leq V \leq 400$ is shown, because~no low-probability candidate regions were found outside this interval.}
    
    \label{fig:eszakProb}
\end{figure}

We next allow the stretching factor $s$ to vary. For~each pair $(r,s)$ we compute the maximum count $K_{\rm obs}(r,s)$ and the probability $p(r,s)$ from the redshift-shuffled catalogs. Figure~\ref{fig:eszakProb} shows the resulting~probabilities.

The main result is that the Northern overdensity remains the strongest feature over a range of spheroidal shapes. Thus, the~signal is not only a consequence of imposing a spherical window. This supports the stability of the Hercules--Corona Borealis Great Wall candidate in the transformed GRB~sample.

The value of $s$ at which the minimum probability occurs gives a simple description of the preferred shape in the transformed coordinates. If~the minimum occurs at $s>1$, the~best counting volume is elongated along the $z$ direction. If~it occurs at $s<1$, the~best counting volume is flattened. If~the minimum remains close to $s=1$, the~spherical description is already close to~optimal.

At the best-fitting pairs $(r,s)$, the~spheroid selects a subset of $K$ GRBs.~The~sky positions of these events overlap the region previously associated with the Hercules--Corona Borealis Great Wall~\citep{hhb14,hbht15,HSZ20}. Since the random catalogs keep the angular positions fixed, the~probability mainly tests whether the observed redshift arrangement creates an unusual three-dimensional concentration for the given sky footprint. 
The most significant Northern-hemisphere candidates for selected spheroidal shapes are listed in Table~\ref{tab:north_best}.

\begin{table}[H]
\centering
\caption{Most significant (<1\%) Northern-hemisphere candidate regions for selected spheroidal shapes. The~probabilities are given in percent. K is the number of GRBs in the~volume.}
\label{tab:north_best}
\begin{tabularx}{\textwidth}{LCCC}
\toprule
\boldmath{$s$} & \textbf{Volume}  & \boldmath{$p$} \textbf{[\%]} & \boldmath{$K$} \\
\midrule
1.13 & 270 & 0.59 & 129  \\
1.13 & 271 & 0.85 & 129  \\
1 & 268 & 0.53 & 125 \\
1 & 269 & 0.77 & 125 \\
1 & 274 & 0.94 & 127 \\
0.91 & 285 & 0.86 & 125 \\
0.91 & 286 & 0.20 & 127 \\
0.91 & 287 & 0.35 & 127 \\
0.91 & 288 & 0.48 & 127 \\
0.91 & 289 & 0.77 & 127 \\
0.91 & 291 & 0.75 & 128 \\
0.91 & 293 & 0.72 & 129 \\
0.72 & 306 & 0.78 & 126 \\
0.72 & 308 & 0.77 & 127 \\
0.72 & 316 & 0.98 & 130 \\
0.59 & 311 & 0.65 & 126 \\
0.59 & 312 & 0.93 & 126 \\
\bottomrule
\end{tabularx}
\end{table}

The key point is the stability of the Northern signal. The~same region remains significant when the counting window is allowed to become non-spherical. This argues against the idea that the signal is only an artefact of using spherical~volumes.

\section{The Southern Hemisphere GRB~Distribution}
\label{sec:south}

The Southern Galactic hemisphere contains 280 GRBs. We apply the same spheroidal search and the same redshift-shuffling test as for the Northern~sample.

\subsection{Recovery of the Spherical~Result}
\label{sec:south_spherical}

For $s=1$, the~result agrees with the spherical analysis of Paper~I. The~Southern hemisphere does not show a big volume overdensity comparable to the Northern one. Paper~I found only one small compact grouping with marginal significance. It consists of four events, with~a nearby fifth event, at~similar redshifts and small angular separations. For~the four-member configuration, the local probability was about $p\simeq 0.007$. 

\subsection{Dependence on Spheroidal~Shape}
\label{sec:south_shape}

\begin{figure}[H]

    \includegraphics[width=0.97\columnwidth]{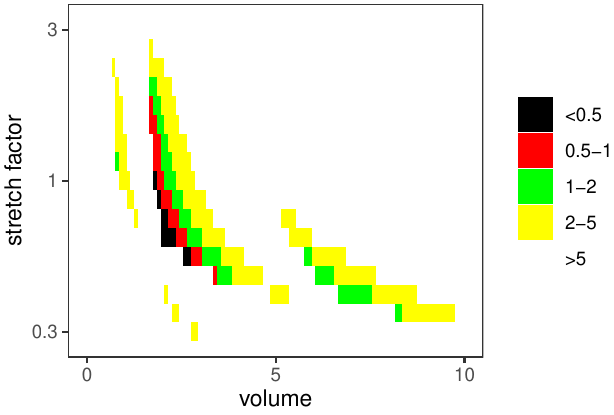}
    \caption{Monte-Carlo
 probabilities for the Southern Galactic hemisphere in the small-volume part of the scan. The~horizontal axis gives the tested volume or scale parameter, and~the vertical axis gives the stretch factor $s$. The~color bar on the right gives the Monte-Carlo probability $p$ in percent. Only the range $V \leq 10$ is shown, because~larger volumes did not produce low-probability candidate regions in this~hemisphere.}
    \label{fig:delProb}
\end{figure}

We then allow $s$ to differ from unity. This tests whether the Southern candidate becomes stronger when the counting volume is elongated or flattened. It also tests whether a new candidate appears for non-spherical windows. For~each value of $s$ we calculate $p_{\min}(s)$ in the same way as for the Northern hemisphere. Figure~\ref{fig:delProb} shows the corresponding Monte-Carlo probabilities for the Southern Galactic~hemisphere.

The spheroidal scan does not reveal any Southern overdensity with a significance comparable to the Northern structure. The~lowest probabilities occur for small groups of events and are sensitive to the exact volume and shape. This behavior is expected for sparse data, where adding or removing one event can change the probability substantially.
\mbox{Tables~\ref{tab:delbestforK} and~\ref{tab:delbest}} summarize these Southern candidates. They show that the lowest probabilities are obtained mainly for $K=4$ and for a limited range of volume--$s$ combinations. 

\begin{table}[H]
\centering
\caption{Most significant Southern-hemisphere candidate regions for different K (number of GRBs). The~probabilities are given in per~cent.}
\label{tab:delbestforK}
\begin{tabularx}{\textwidth}{lCCCC}
\toprule
\boldmath{$s$} & \textbf{Volume}  & \boldmath{$p$} \textbf{[\%]} & \boldmath{$K$} & \textbf{Comment} \\
\midrule
1.3 & 0.2 & 6.8 & 2 & not significant \\
1.15 & 0.9 & 1.9 & 3 &  \\
0.63 & 2.1 & 0.2 & 4 &  \\
0.41 & 6.8 & 1.07 & 5 &  \\
0.74 & 7.8 & 5.4 & 6 & not significant \\
\bottomrule
\end{tabularx}
\end{table}
\unskip

\begin{table}[H]
\centering
\caption{Most significant Southern-hemisphere candidate regions for $K$ (number of GRBs) $=4$, where the probabilities ($p$) are smaller than 2 percent. The~entries give $p$ [\%].}
\label{tab:delbest}
\small
\setlength{\tabcolsep}{3pt}
\begin{tabularx}{\textwidth}{lCCCCCCCCCCC}
\toprule
\textbf{Volume}\boldmath{$\backslash s$} & \boldmath{$2$} & \boldmath{$1.75$} & \boldmath{$1.5$} & \boldmath{$1.3$} & \boldmath{$1.15$} & \boldmath{$1$} & \boldmath{$0.86$} & \boldmath{$0.74$} & \boldmath{$0.63$} & \boldmath{$0.55$} & \boldmath{$0.47$ }\\
\midrule
1.8 & 1.29 & 0.93 & 0.66 & -- & -- & -- & -- & -- & -- & -- & -- \\
1.9 & 1.78 & 1.29 & 0.91 & 0.71 & 0.58 & 0.45 & -- & -- & -- & -- & -- \\
2.0 & -- & 1.72 & 1.29 & 0.98 & 0.78 & 0.66 & 0.37 & -- & -- & -- & -- \\
2.1 & -- & -- & 1.71 & 1.28 & 1.04 & 0.82 & 0.52 & 0.30 & \textbf{0.2} & -- & -- \\
2.2 & -- & -- & -- & 1.66 & 1.32 & 1.11 & 0.65 & 0.44 & 0.26 & -- & -- \\
2.3 & -- & -- & -- & -- & 1.69 & 1.39 & 0.84 & 0.56 & 0.35 & -- & -- \\
2.4 & -- & -- & -- & -- & -- & 1.73 & 1.05 & 0.7 & 0.45 & -- & -- \\
2.5 & -- & -- & -- & -- & -- & -- & 1.34 & 0.92 & 0.58 & -- & -- \\
2.6 & -- & -- & -- & -- & -- & -- & 1.67 & 1.15 & 0.78 & -- & -- \\
2.7 & -- & -- & -- & -- & -- & -- & -- & 1.37 & 0.92 & 0.39 & -- \\
2.8 & -- & -- & -- & -- & -- & -- & -- & 1.66 & 1.16 & 0.48 & -- \\
2.9 & -- & -- & -- & -- & -- & -- & -- & -- & 1.35 & 0.59 & -- \\
3.0 & -- & -- & -- & -- & -- & -- & -- & -- & 1.54 & 0.74 & -- \\
3.1 & -- & -- & -- & -- & -- & -- & -- & -- & 1.81 & 0.92 & -- \\
3.2 & -- & -- & -- & -- & -- & -- & -- & -- & -- & 1.06 & -- \\
3.3 & -- & -- & -- & -- & -- & -- & -- & -- & -- & 1.27 & -- \\
3.4 & -- & -- & -- & -- & -- & -- & -- & -- & -- & 1.43 & -- \\
3.5 & -- & -- & -- & -- & -- & -- & -- & -- & -- & 1.68 & 0.99 \\
3.6 & -- & -- & -- & -- & -- & -- & -- & -- & -- & -- & 1.17 \\
3.7 & -- & -- & -- & -- & -- & -- & -- & -- & -- & -- & 1.35 \\
3.8 & -- & -- & -- & -- & -- & -- & -- & -- & -- & -- & 1.54 \\
3.9 & -- & -- & -- & -- & -- & -- & -- & -- & -- & -- & 1.78 \\
\bottomrule
\end{tabularx}
\end{table}
\unskip

\subsection{Candidate Grouping and~Robustness}
\label{sec:south_candidate}

The best Southern candidate contains only a small number of GRBs. Its probability is therefore strongly affected by discreteness and by the look-elsewhere effect. For~this reason, we treat it only as a provisional~candidate.

Two of the events, GRB 050822 and GRB 050318, have a small angular separation of $1.03^{\circ}$ and similar spectroscopic redshifts, $z=1.434$ and $z=1.4436$. They were also detected within a five-month interval. This makes the pair noteworthy. However, the~angular separation is much larger than the scale expected for strong gravitational lensing. The~couple therefore cannot be explained as a simple lensing duplicate. More data are needed to decide whether the group is part of a real large-scale structure or a chance alignment in a sparse~sample.

The Southern result should be interpreted with caution. The~random catalogs preserve the observed angular distribution and the global redshift distribution, but~they do not model the full selection function of GRB redshift measurements. A~real confirmation would require independent tracers, such as galaxies, quasars, or~star-forming galaxies, in~the same three-dimensional~region.

\section{Discussion}
\label{sec:discussion}

The main result of this analysis is that the Northern GRB overdensity found in Paper~I remains present when the search window is changed from a sphere to a spheroid. This is an important check. It shows that the signal is not tied to one fixed window shape. The~overdensity is also found in the region associated with the Hercules--Corona Borealis Great Wall. We therefore describe the result as support for the stability of that candidate structure in the GRB redshift~sample.

The result does not, by~itself, prove that the Hercules--Corona Borealis Great Wall is a physical matter structure. GRBs are rare events, and~only a fraction of them have spectroscopic redshifts. The~redshift sample is affected by satellite exposure, afterglow brightness, dust extinction, and~the availability of follow-up spectroscopy. The~redshift-shuffling test controls for the observed sky positions and the observed redshift distribution, but~it cannot remove every possible selection~effect.

The Southern Galactic hemisphere behaves differently. The~spheroidal scan does not find a structure with a significance similar to the Northern one. A~small candidate grouping is present, but~it contains only a few events and is sensitive to the details of the scan. This supports the view that the Southern sample is broadly consistent with a homogeneous distribution within the limits of the present~data.

The comparison between the two hemispheres is suggestive, but~it should not be over-interpreted. It shows that the Northern overdensity is unusual within this GRB sample. It does not automatically imply a global violation of the Cosmological Principle. GRBs trace special astrophysical environments, especially massive star formation for long-duration events, and~their occurrence rate is not a direct linear measure of the total matter density.
The cosmological interpretation is also sensitive to the assumed background model: in a flat dust + ($\Lambda $)~\citep{paal92,1998AJ....116.1009R,1999ApJ...517..565P} cosmology, where a positive cosmological constant corresponds to a vacuum-energy, or~($\Lambda $)-type dark-energy, component, the~inferred comoving scale and phase coherence of apparent galaxy-redshift regularities can differ from those in a dust-only~model.

The astrophysical interpretation is therefore limited. A~local increase in the GRB rate may be caused by a large-scale matter overdensity, but~it may also be affected by the environments that favor massive star formation. Galaxy interactions and mergers can increase star formation and may increase the GRB rate in selected regions~\citep{BalazsRing2015}. For~this reason, a~GRB overdensity should be seen as a candidate structure until it is checked with independent~tracers.

Similar caveats were discussed by \citet{HSZ20}, who considered astrophysical explanations and observational-bias effects alongside cosmological interpretations of the Hercules--Corona Borealis Great Wall. This supports treating a GRB overdensity as a candidate tracer of large-scale structure rather than as a direct linear map of the total matter~density.

Future work should focus on two points. First, the~analysis should be repeated as the GRB spectroscopic sample grows. Second, the~same comoving volumes should be tested with denser tracers, such as galaxies, quasars, and~emission-line galaxies. Next-generation spectroscopic surveys and early data releases, including DESI and Euclid, provide natural routes for such cross-checks~\citep{DESI2025JCAP021,EuclidQ1Spec2025}. Such tests are needed to decide whether the GRB overdensity is part of the underlying matter distribution or is mainly a feature of the GRB selection and formation~process.

\section{Summary}
\label{sec:summary}

We have re-analyzed the three-dimensional distribution of 542 GRBs with spectroscopic redshifts using axisymmetric spheroidal counting volumes rather than the spherical volumes used in Paper~I~\cite{horv26MNRAS}. The~angular positions of the observed GRBs are kept fixed in the Monte-Carlo samples, while the redshifts are shuffled within each Galactic~hemisphere.

In the Northern Galactic hemisphere, the~overdensity associated with the Hercules--Corona Borealis Great Wall remains significant for a range of spheroidal shapes. This supports the stability of the previously reported signal and shows that it is not caused only by the use of spherical counting~volumes.

In the Southern Galactic hemisphere, the~same method does not find a structure of comparable significance. A~small candidate grouping is present, but~it  contains too few events for a strong~conclusion.

These results show that spheroidal resampling is a useful test for GRB overdensity searches. At~the same time, the~limits of the GRB redshift sample remain important. Independent confirmation with denser tracers in the same comoving volume is needed before any GRB candidate structure can be treated as a confirmed large-scale matter~structure.

\vspace{6pt} 



\authorcontributions{Conceptualization, {I.H., Z.B., L.G.B., J.H. (Jon Hakkila), B.K., I.I.R., P.V. and~S.P.}; 
data curation, {I.H., Z.B., L.G.B., J.H. (Jon Hakkila), B.K., I.I.R., P.V. and~S.P.}; formal analysis, I.H. and Z.B.; investigation, all; methodology, {I.H., Z.B., L.G.B., J.H. (Jon Hakkila), B.K., I.I.R., P.V. and~S.P.}; project administration, S.P. and I.H.; resources, {I.H., Z.B., L.G.B., J.H. (Jon Hakkila), B.K., I.I.R., P.V. and~S.P.}; software, I.H. and Z.B.; supervision, {I.H., Z.B., L.G.B., J.H. (Jon Hakkila), B.K., I.I.R., P.V. and~S.P.}; validation, {I.H., Z.B., L.G.B., J.H. (Jon Hakkila), B.K., I.I.R., P.V. and~S.P.}; visualization, {J.H. (Janos Horvath), S.P., Z.B., L.G.B.,
 and~I.H.}; writing---original draft, {I.H., Z.B., L.G.B., J.H. (Jon Hakkila), B.K., I.I.R., P.V. and~S.P.}; and writing---review and editing, {I.H., Z.B., L.G.B., J.H. (Jon Hakkila), B.K., I.I.R., P.V. and~S.P.} All authors have read and agreed to the published version of the manuscript.}

\funding{This research received no external funding.}

\dataavailability{The data underlying this paper are available in Gamma-Ray Burst Online Index (GRBOX) database published by the Caltech Astronomy Department
 (\url{https://sites.astro.caltech.edu/grbox/grbox.php} accessed on 27.08.2026), Jochen Greiner's table  (\url{https://www.mpe.mpg.de/~jcg/grbgen.html} accessed on 27.08.2026), and relevant  Gamma-ray
 Coordinates Network  (\url{https://gcn.gsfc.nasa.gov/gcn3_archive.html} accessed on 27.08.2026) messages. The~code and data used for this study are available at \url{https://github.com/horvathist/GRBSpherical2025} accessed on 27.08.2026.
}


\conflictsofinterest{Author Janos. Horvath was employed by the company Visionary Tech \& Event Solutions. The~remaining authors declare that the research was conducted in the absence of any commercial or financial relationships that could be construed as potential conflicts of~interest.} 




\begin{adjustwidth}{-\extralength}{0cm}

\reftitle{References}

\PublishersNote{}
\end{adjustwidth}
\end{document}